\pdfoutput=1
\documentclass{pspum-l}

\usepackage{graphicx}
\usepackage{amsmath}
\usepackage{amssymb}
\usepackage{tikz,pgfplots}

\theoremstyle{definition}

\theoremstyle{remark}

\DeclareMathOperator{\Tr}{Tr}
\DeclareMathOperator{\Spec}{Spec}

\newcommand{\C}{\mathbb{C}}

\begin{document}

\title{Superconformal field theories and cyclic homology}


\author{Richard Eager}
\address{Department of Physics, McGill University Montr\'eal, QC, Canada \\
}
\curraddr{Mathematisches Institut, Universit\"at Heidelberg, Heidelberg, Germany}
\email{reager@physics.mcgill.ca}



\date{}

\begin{abstract}
One of the predictions of the AdS/CFT correspondence is the matching of protected operators between a superconformal field theory and its holographic dual.  We review the spectrum of protected operators in quiver gauge theories that flow to superconformal field theories at low energies.  The spectrum is determined by the cyclic homology of an algebra associated to the quiver gauge theory.  Identifying the spectrum of operators with cyclic homology allows us to apply the Hochschild-Kostant-Rosenberg theorem to relate the cyclic homology groups to deRham cohomology groups.  The map from cyclic homology to deRham cohomology can be viewed as a mathematical avatar of the passage from open to closed strings under the AdS/CFT correspondence.
\end{abstract}

\maketitle

\section{Introduction}
One of the most basic predictions of the AdS/CFT correspondence is the equivalence between the operators in a conformal field theory and its dual gravitational theory.  Since a conformal field theory is determined by its collection of operators and their correlation functions, the matching of operators and their scaling dimensions on both sides of the duality is a natural starting point to understand the duality.  However the spectrum of operators can vary with the coupling and it is often impossible to determine the scaling dimensions of operators at strong coupling without resorting to the AdS/CFT correspondence.  For supersymmetric gauge theories that flow to $\mathcal{N} = 1$ superconformal field theories (SCFTs) at low energies, there is a class of BPS operators with protected scaling dimensions.  We review how these protected operators can be determined using $\mathcal{Q}$-cohomolgy or equivalently cyclic homology.

The key new idea is re-writing the $\mathcal{Q}$-cohomology groups in terms of cyclic homology.  This allows many techniques and theorem developed in mathematics to be applied to the analysis of quiver gauge theories.  In particular for quiver gauge theories that are dual to type IIB string theory on $AdS_5$ times a Sasaki-Einstein manifold, the equality of the gauge theory and gravity superconformal indices was proved in \cite{Eager:2012hx} using the Hochschild-Kostant-Rosenberg (HKR) theorem relating cyclic homology of an algebra associated to the quiver gauge theory to the deRham cohomology of the Calabi-Yau cone over the Sasaki-Einstein manifold.

Quiver gauge theories dual to Sasaki-Einstein manifolds have been intensely studied, but they are at very special points in the moduli space of the SCFT that they flow to at low energy.  In this paper we take the first steps to apply cyclic homology to determine the spectrum of general quiver gauge theories.  While the superconformal index is constant over the SCFT moduli space, the indivual $\mathcal{Q}$-cohomolgy groups can vary over the moduli space.  We illustrate this phenomena with the $\beta$-deformation of $\mathcal{N}=4$ supersymmetric Yang-Mills in section \ref{sec:beta}.  In this case, the HKR theorem no longer applies, but we expect the matching of operators in the AdS/CFT correspondence can still be viewed as the passage from cyclic homology to Poisson homology.

This paper is based on a talk the author gave at String-Math 2014 and reviews joint work with J. Schmude and Y. Tachikawa that appeared in \cite{Eager:2012hx}.  Section \ref{sec:2} reviews the identification of protected BPS operators with elements of cyclic homology.  Section \ref{sec:examples} applies these techniques to two familiar examples.  The first example is $\mathcal{N} = 4$ supersymmetric Yang-Mills theory and is covered in section \ref{sec:sym}.
The second example in section \ref{sec:beta} is a new application of cylic homology to analyze the $\beta$-deformation of $\mathcal{N}=4$ supersymmetric Yang-Mills, extending the original analysis in \cite{Berenstein:2000ux}.  Cyclic homology can also be used to analyze the other marginal and relevant deformations of $\mathcal{N}=4$ supersymmetric Yang-Mills theory \cite{Leigh:1995ep, Berenstein:2000ux}.  Some preliminary results in this direction will be briefly discussed in the final section and are part of work-in-progress.
\section{Cyclic Homology of Quiver Gauge Theories}
\label{sec:2}
\subsection{BPS Operators in SCFTs}
\begin{table}
\centering
\begin{tabular}{|c|c|c|}
\hline
Letter & $(j_1, j_2)$ & $\mathcal{I}$  \\ \hline\hline
$\phi$ & $(0,0)$ & $t^{3r}$ \\ \hline
$\overline{\psi}_{\dot{2}} $& $(0,1/2)$ & $-t^{3(2-r)}$ \\ \hline
\multicolumn{3}{c}{}\\
\hline
\hline
$\partial_{\pm -}$ & $(\pm 1/2, 1/2)$ & $t^3 y^{\pm 1}$ \\
\hline
\end{tabular}
\quad
\quad
\begin{tabular}{|c|c|c|}
\hline
Letter & $(j_1, j_2)$ & $\mathcal{I}$  \\ \hline\hline
$\lambda_1$ & $(1/2,0)$ & $-t^{3} y$ \\ \hline
$\lambda_2 $& $(-1/2, 0)$ & $-t^{3} y^{-1}$ \\ \hline
$\overline{f}_{\dot{2}\dot{2}}$ & (0,1) & $t^6$ \\ \hline
\hline
$\partial_{\pm -}$ & $(\pm 1/2, 1/2)$ & $t^3 y^{\pm 1}$ \\
\hline
\end{tabular}
\caption[tab1]{Fields contributing to the $\mathcal{Q}$-cohomology groups, from a chiral multiplet (left) and from a vector multiplet (right).  The equation of motion $\partial_{22} \lambda_{1} = \partial_{12} \lambda_{2}$ must also be accounted for \cite{Dolan:2008qi, Gadde:2010en}. }
\label{tab:let}
\end{table}
We consider $\mathcal{N} = 1$ gauge theories on $S^1 \times S^3$ with a Lagrangian description that flow to $\mathcal{N}=1$ SCFTs at low energies.  In these theories, a special role is played by BPS operators, which are annihilated by one of the supercharges $\mathcal{Q} = \overline{Q}_{\dot{1}}$.
The quantum numbers of BPS operators saturate the following inequality
\begin{equation*}
\{\mathcal{Q}, \mathcal{Q}^{\dagger} \} = E - 2 j_2 - \frac{3}{2} r \ge 0,
\end{equation*}
where $E$ is the energy, $r$ is the R-charge, and $(j_1,j_2)$ are the Lorentz spins.
This is the unitary bound for the $\mathcal{N} = 1$ superconformal algebra.
States saturating this bound are called BPS.
Our goal is to compute the $\mathcal{Q}$-cohomology for quiver gauge theories in the large $N$ limit.
The superconformal primaries of the $\mathcal{N} = 1$ superconformal algebra are by definition annihilated by $\mathcal{Q}^{\dagger}$.  Since states in $\mathcal{Q}$-cohomology are $\mathcal{Q}$-closed, these states satisfy the unitary bound.  In the quiver gauge theories that we are considering, BPS operators are formed by traces of products of letters that satisfy the bound $E - 2 j_2 - \frac{3}{2} r \ge 0$.  We list these letters in Table \ref{tab:let}.

The $\mathcal{Q}$-cohomology groups are labeled by the $SU(2)_r$ spin $j_2$.  The zeroth cohomology group consists of the elements of the chiral ring \cite{Romelsberger:2005eg}.  The higher cohomology groups correspond to short representations of the superconformal algebra known as $1/4$ BPS operators\footnote{These are $1/16$ BPS operators in $\mathcal{N} = 4$ SYM.}.  Detailed information about these short representations can be found in \cite{Dolan:2008qi, Gadde:2010en}.

The single-trace index is the Euler characteristic of these cohomology groups, and it is defined by
$$\mathcal{I}_{s.t.} = \Tr_{single-trace \; op.} (-1)^F t^{2(E+j_2)} y^{2 j_1}.$$
From the single-trace index, the full superconformal index can be determined  \cite{Romelsberger:2005eg,Kinney:2005ej}.
In the large-$N$ limit, the contribution to the index from the letters $\partial_{\alpha}$ and $\lambda_{\alpha}$ cancels \cite{Eager:2012hx}.  We therefore restrict our attention to the operators constructed from gauge invariant operators formed from words in the letters $\phi, \overline{\psi}_2,$ and $\overline{f}_{22}$.  Since the superconformal index is a more robust and easier to calculate quantity than the individual $\mathcal{Q}$-cohomology groups, it serves as an important guide and check on our understanding of $\mathcal{Q}$-cohomology.

While the index is invariant under exactly marginal deformations, the individual $\mathcal{Q}$-cohomology groups can jump.  As the values of the coupling constants approach special loci in moduli space, the anomalous dimension of a long multiplet can decrease until it saturates a BPS bound.  At this loci in moduli space the long multiplet can decompose into a direct sum of short multiplets.  The short multiplets will contribute to the $\mathcal{Q}$-cohomology groups, so the $\mathcal{Q}$-cohomology will generically jump at the special loci.  However, the index is constructed so that the contributions to the index from short multiplets that can form a long multiplet cancel.  Therefore, the index is invariant under exactly marginal deformations.  We will see an explicit example of this phenomena when we examine the $\beta$-deformation in section \ref{sec:beta}.  Another important example is the change in the $\mathcal{Q}$-cohomology groups when a gauge coupling vanishes.

\subsection{Quiver Gauge Theories}
A gauge theory consists of a gauge group $G$ along with matter fields in a representation $V$ of $G$ and interactions encoded by a Lagrangian.  Theories with
$\mathcal{N} = 1$ supersymmetry can be efficiently described using superspace.  A large class of $\mathcal{N}=1$ supersymmetric gauge theories can be specified by a quiver and a superpotential.  A quiver $Q = (Q_0, Q_1, h, t)$ is a collection of vertices $Q_0$ and arrows $Q_1$ along with maps $h, t: Q_1 \rightarrow Q_0$ which specify the head and tail of an arrow.  Given a quiver $Q$ and superpotential $W$, we can define a gauge theory by setting the gauge group to be 
$$G = \prod_{v \in Q_0} U(N_v).$$  To each arrow $a \in Q_1$ we assign a chiral superfield $\Phi_a$  which transforms in the fundamental representation of $U(N_{h(a)})$ and the anti-fundamental representation of $U(N_{t(a)})$.  If $a$ is a closed loop, then the superfield transforms in the adjoint representation of $U(N_{h(a)})$.  Finally the superpotential $W$ is a sum of gauge invariant operators.  Gauge invariance requires that the superpotential $W$ is a linear combination of cyclic words in the quiver.  A fancier way of writing this is $W \in \C Q_{cyc} = \C Q/[\C Q, \C Q]$. From a quiver with potential $(Q,W)$ we can construct its superpotential algebra $\mathcal{A}_{Q,W} = \C Q/(\partial W).$ 

The action of the supercharge $\mathcal{Q}$ on a quiver gauge theory is 
\begin{align*}
\mathcal{Q} \phi_e 					& = 0 \\
\mathcal{Q} \overline{\psi}_{e,\dot{2}}	& = \frac{\partial W(\phi)}{\partial \phi_e} \\
\mathcal{Q} \overline{f}_{v, \dot{2} \dot{2}} &= \sum_{h(e) = v} \phi_e \overline{\psi}_{e,\dot{2}} - \sum_{t(e) = v} \overline{\psi}_{e, \dot{2}} \phi_e.
\end{align*}
A direct calculation shows that $\mathcal{Q}$ is nilpotent
\begin{align*}
\mathcal{Q}^2 \overline{f}_{v, \dot{2} \dot{2}} & =  \sum_{h(e) = v} \phi_e \frac{\partial W(\phi)}{\partial \phi_e} - \sum_{t(e) = v} \frac{\partial W(\phi)}{\partial \phi_e}\phi_e \\
& = 0.
\end{align*}
While physically obvious, this relationship is a syzygy in the superpotential algebra $\mathcal{A}$ \cite{MR1247289}.  Thus we have uncovered a direct physical explanation for the appearance of this syzygy in the pioneering definition of Calabi-Yau algebras in \cite{Berenstein:2002fi}.
\subsection{Ginzburg's DG Algebra}
\begin{table}
\centering
\begin{tabular}{|c|c|c|}
\hline
Letter & Ginzburg DG & grading \\ \hline \hline
$\phi$ & $x_{e}$ & 0  \\ \hline
$\overline{\psi}_2$ & $x_{e}^{*}$ & 1  \\ \hline
$\overline{f}_{22}$ & $t_v$ & 2\\
\hline
\hline
$\overline{Q}_{\dot{1}}$ & $d$ & $-1$ \\ 
\hline
\end{tabular}
\caption[tab2]{The dictionary between physical fields and generators of Ginzburg's DG algebra.}
\label{tab:ginzburg}
\end{table}
The algebra generated by the letters $\phi_e,  \overline{\psi}_{e,\dot{2}}$, and $\overline{f}_{v, \dot{2} \dot{2}}$ with the differential $\mathcal{Q}$ is known as Ginzburg's DG algebra \cite{Ginzburg:2006fu}.  In the DG algebra, the generators are denoted by $x_e, x^{*}_e$, and $t_v$ and the differential is denoted by $d.$  This dictionary is shown in Table \ref{tab:ginzburg}.  To the quiver $Q$ we associate the quiver $\widehat{Q}$ with the same vertices as $Q$ in the following way.  For each arrow $e \in Q$ there is a corresponding arrow $x_e$ in $\widehat{Q}$ and an arrow $x^{*}_e$ with the opposite orientation.  For each vertex $v$ of $Q$ there is a loop $t_v$ based at vertex $v$ in $\widehat{Q}.$
The grading is simply twice the $j_2$ spin.  In particular the letters $x_e, x^{*}_e$, and $t_v$ are assigned charges 0,1, and 2 under the grading.  The differential $d$ has charge $-1$ under the grading.
Let $\mathfrak{D} = \mathbb{C}\langle x_e, x^{*}_e, t_v \rangle$ be the free DG algebra generated by the paths $x_e, x^{*}_e$, and $t_v.$   The degree zero homology is $HH_0(\mathfrak{D}) =  \mathbb{C}\langle x_e \rangle / \left( \partial W(x) / \partial x_e \right) .$  Thus, the degree zero homology $HH_0(\mathfrak{D})$ is isomorphic to the superpotential algebra $\mathcal{A}.$  If the algebra $\mathcal{A}$ is a Calabi-Yau algebra of dimension three, then all positive degree homology groups $HH_{> 0}(\mathfrak{D})$ vanish \cite{Ginzburg:2006fu}.  In particular, there is a quasi-isomorphism of chain complexes $\mathfrak{D} \twoheadrightarrow \mathcal{A}.$  Here $\mathcal{A}$ represents the complex with the algebra $\mathcal{A}$ in degree zero and zero in all other degrees.
\subsection{Cyclic Homology}
\label{sec:cyc}
Let  $[\mathfrak{D}, \mathfrak{D} ]$ be the $\mathbb{C}$-linear space spanned by the commutators of elements in $\mathfrak{D}.$  Then $\mathfrak{D}_{cyc} := \mathfrak{D}/(\mathbb{C} + [\mathfrak{D}, \mathfrak{D} ])$ is the space of cyclic words in $\mathfrak{D}$ up to scalar multiplication, or in other words, $\mathfrak{D}_{cyc} $ is the $\mathbb{C}$-vector space generated by the space of paths in $\widehat{Q}.$  Equivalently, $\mathfrak{D}_{cyc}$ is the $\mathbb{C}$-vector space
of single-trace operators formed by $\phi_e, \overline{\psi}_{e,\dot{2}},$ and $ \overline{f}_{v, \dot{2} \dot{2}}$.
For large-N quiver gauge theories with unitary groups, the gauge-invariant operators correspond to the closed paths in the Ginzburg quiver $\widehat{Q}.$  The BPS operators are then elements of the homology groups $H_{\bullet}(\mathfrak{D}_{cyc}, d)$ where the differential $d$ in the DG algebra is the SUSY differential $\mathcal{Q}.$ If $\mathcal{A}$ is a Calabi-Yau algebra of dimension three, then since 
$\mathfrak{D} \twoheadrightarrow \mathcal{A}$ is a quasi-isomorphism there is an isomorphism \cite{MR923137}
$$\overline{HC}_{\bullet}(\mathcal{A}) \cong H_{\bullet}(\mathfrak{D}_{cyc}, d).$$
Thus the physical $\mathcal{Q}$-cohomology obtained from the homology of Ginzburg's DG algebra is isomorphic to the reduced (negative) cyclic homology of the superpotential algebra $\mathcal{A}$. 

As an example, we consider the spin zero chiral primary operators corresponding to elements of the zeroth cyclic homology group $\overline{HC}_{0}(\mathcal{A}) \cong \mathcal{A}/ [ \mathcal{A}, \mathcal{A} ] \cong \mathcal{A}_{cyc}.$
These operators can be easily understood as follows \cite{Berenstein:2000hy,Berenstein:2000ux}.  To each element $a$ of the path algebra $\mathcal{A}$, a representation associates an operator $\mathcal{O}(a)$.  By the cyclic property of the trace
$$\mathcal{O}(ab) = \Tr(ab) = \Tr(ba) = \mathcal{O}(ba),$$
so the representation must factor through the map
$$\mathcal{A} \rightarrow \mathcal{A}/ [ \mathcal{A}, \mathcal{A} ] \cong \mathcal{A}_{cyc}.$$
In other words, if the F-term equations imply that $\Tr(ab) = \gamma \Tr(ba)$ for a constant $\gamma \neq 1,$ then the operator $\mathcal{O}(ab)$ must be $\mathcal{Q}$-exact.  Therefore the operator cannot be a chiral primary.  We will use this argument to find the chiral primaries in the $\beta$-deformation in section \ref{sec:beta}.
\section{Examples}
\label{sec:examples}
\subsection{$\mathcal{N} = 4$ Super Yang-Mills}
\label{sec:sym}
In this section we determine the spectrum of protected operators in $\mathcal{N} = 4$ super Yang-Mills theory with respect to a $\mathcal{N} = 1$ superconformal subgroup of the $\mathcal{N} = 4$ superconformal group.
The matter content of the $\mathcal{N} = 4$ super Yang-Mills theory consists of three $\mathcal{N} = 1$ adjoint chiral multiplets $x,y,z$ and a vector multiplet.  The superpotential is
$W = xyz - xzy.$  The F-term relations imply that the three variables $x,y$, and $z$ commute and so the superpotential algebra 
$$\mathcal{A} \cong \C \langle x,y,z \rangle/ \langle xy - yx, yz-zy,zx-xz\rangle  \cong \mathbb{C}[x,y,z] \cong A$$ is commutative.  In this example, $X = \Spec A$ can be thought of a the variety $\mathbb{C}^3.$
 
The Hochschild homology of a commutative algebra $A$ is isomorphic to the space of algebraic differential forms,
$$\Omega^n_{A} \cong HH_n(A).$$
In this example, $HH_0(A)$ is spanned by the polynomials in $x,y,z$ and $HH_1(A)$ is spanned by the differential forms $f dx + g dy + h dz$, where $f,g,h$ are polynomials in $x,y,z.$
This follows from the Hochschild-Kostant-Rosenberg theorem.
Using the relation between Hochschild and cyclic homology groups \cite{Eager:2012hx}
\begin{align*}
HH_0&=HC_0, \\
HH_1&=HC_0\oplus HC_1,\\
HH_2&=HC_1\oplus HC_2,\\
HH_3&=\overline{HC}_2,
\end{align*}
we determine the cyclic homology groups of $A$.
\begin{table}[htdp]
\begin{center}
\begin{tabular}{|c|c|c|c|c|c|c|c|c|}
\hline
 & $1$ & $t^2$ & $t^4$ & $t^6$ & $t^8$ & $t^{10}$ & $t^{12}$ & $\dots$ \\
\hline
$HC_{0}$ & 1 & 3 & 6 & 10 & 15 & 21 & 28 & $\dots$ \\
$HC_{1}$ & 0 & 0 & 3 & 8 & 15 & 24 & 35 & $\dots$ \\
$HC_{2}$ & 0 & 0 & 0 & 1 & 3 & 6& 10 & $\dots$ \\
\hline
$\mathcal{I}_{s.t}(t)$ & 1 & 3 & 3 & 3 & 3 & 3 & 3 & $\dots$ \\
\hline
\end{tabular}
\caption{Cyclic homology group dimensions for $\mathcal{N}=4$ SYM}
\label{tab:1}
\end{center}
\label{default}
\end{table}%
The cyclic homology groups and their contribution to the single-trace index are displayed in table \ref{tab:1}.  The states reproduce the supergravity states contributing to the superconformal index \cite{Gunaydin:1984fk, Kinney:2005ej, Eager:2012hx}.  Note that all higher cyclic homology groups vanish.  This corresponds to the absence of higher-spin particles in the gravity theory.

As an example we explain the determination of $\overline{HC}_{0}(\mathcal{A})$ in more detail.  Following the original discussion in \cite{Witten:1998qj}, we consider the operators
$$\mathcal{O} =  \Tr \Phi^{z_1} \Phi^{z_2} \dots \Phi^{z_k}$$
where $\Phi$ is an $x,y,$ or $z$ field.
If the operator is symmetric in its indices then it is in a short representation.  If not, then (part of) the operator is a descendant.
The descendants all arise from F-term equations, which is equivalent to being $\mathcal{Q}$-exact or $d$-exact in Ginzburg's DG algebra.  Therefore, the degree 0 piece of $H_{\bullet}(\mathfrak{D}_{cyc}, d)$ consists of the chiral primary operators.

\begin{figure}[htbp]
\begin{center}
\includegraphics[width=8cm]{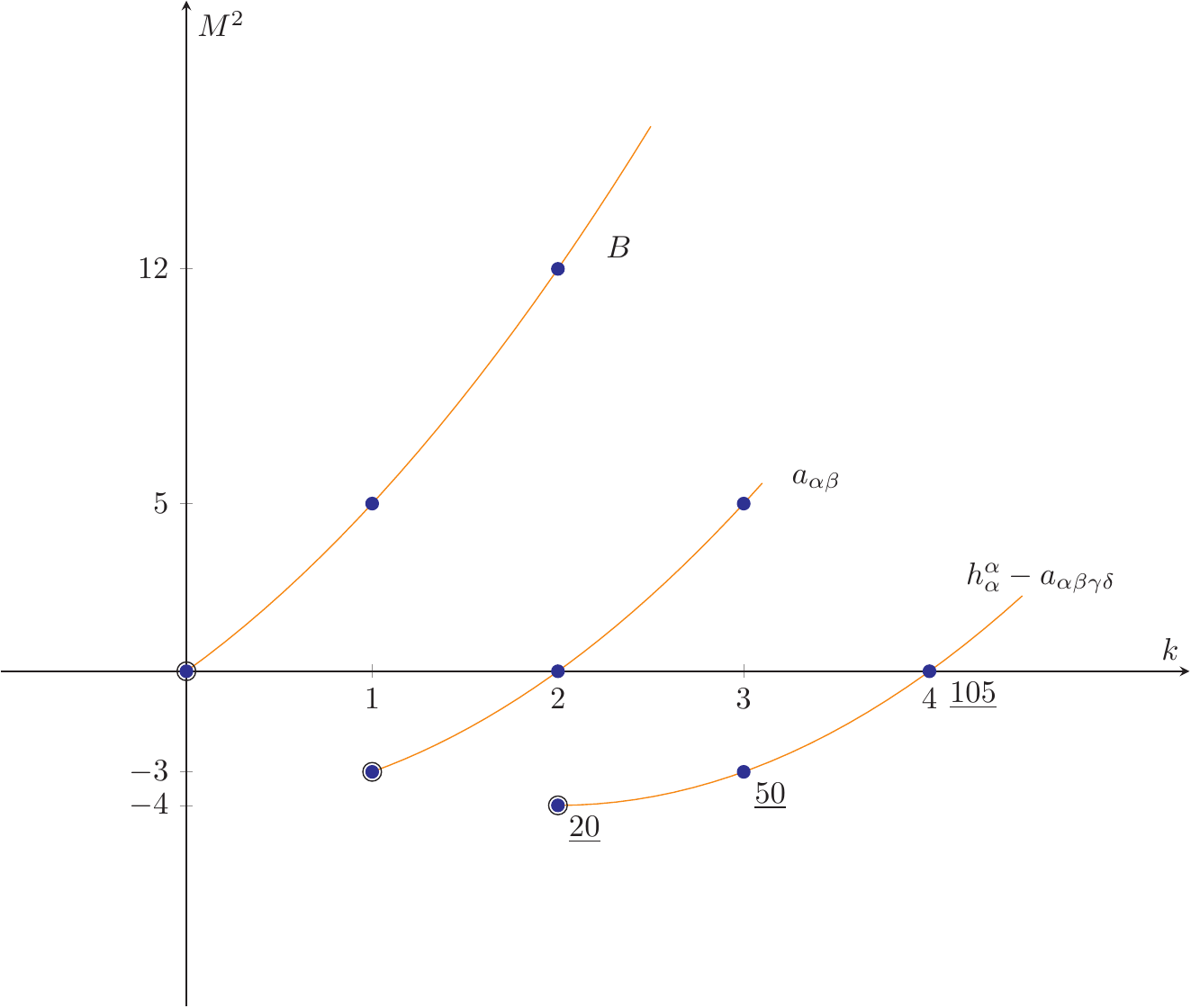}
\caption{Kaluza-Klein mass spectrum of low-lying scalar excitations on $AdS_5 \times S^5$ \cite{Kim:1985ez}.}
\end{center}
\label{fig:KK}
\end{figure}

In this example it is possible to match all of the protected operators between the gauge theory and the protected Kaluza-Klein operators in type IIB supergravity on $AdS_5 \times S^5.$  Of course this is much easier using $\mathcal{N} = 4$ superconformal representations, but we will illustrate the method only using a $\mathcal{N} = 1$ subalgebra. 
The operator $\mathcal{O}$ has conformal dimension $k$ and transforms in the $k$-th symmetric representation, $\text{Sym}^k \bold{3}$, of $SU(3)$.  By the usual AdS/CFT relation, this operator is dual to a supergravity state of spin zero and mass
$$m^2 = k(k-4).$$
These operators precisely correspond to the Kaluza-Klein modes labeled by $h_{\alpha}^{\alpha} - a_{\alpha \beta \gamma \delta}$ originally determined in  \cite{Kim:1985ez} and reproduced in figure \ref{fig:KK}.  There, the operators $\bold{20}, \bold{50}, \bold{105}$ are in the symmetric-traceless representations of $SO(6).$  However, only the operators in the symmetric representations of $SU(3)$ are primaries with respect to a fixed $\mathcal{N} = 1$ subalgebra of the $\mathcal{N}=4$ superconformal group.

In the examples in  \cite{Eager:2012hx} where $X$ is a cone over a Sasaki-Einstein manifold, the algebra $\mathcal{A}$ will have the same Hochschild and cyclic homology as a (commutative) variety $X$.  Even though $\mathcal{A}$ and $A$ are not isomorphic in this case, our general strategy can still be applied.  We can again use the equality of the cyclic homology of $\mathcal{A}$ and $X$ to related the spectrum of protected operators in the gauge theory described by $\mathcal{A}$ to the protected spectrum of Kaluza-Klein modes on $X$. 
\subsection{$\beta$-deformation}
\label{sec:beta}
We now determine the spectrum of protected operators in the $\beta$-deformation of $\mathcal{N} = 4$ super Yang-Mills theory.  
The $\beta$-deformation is a quiver gauge theory with potential $W =  q xyz - q^{-1} xzy$ where $q = e^{i \beta}.$
The F-term relations are 
\begin{align*}
xy & = q^{-2} yx \\
yz & = q^{-2} zy \\
zx & = q^{-2} xz  
\end{align*}
which shows that the matrices parametrizing the moduli space of vacua are non-commutative.  The cyclic homology groups were computed in \cite{MR1252939}.  From the graded components of the cylic homology group, we extract the number of BPS operators with fixed $j_2$ spin and their contribution to the index.  The operators and their contribution to the index are listed in table \ref{tab:beta}.  We see that while the individual cohomology groups differ from those in $\mathcal{N} = 4$ super Yang-Mills, the index is the same.  This is as expected, since the index is invariant under exactly marginal deformations.
\begin{table}[htdp]
\begin{center}
\begin{tabular}{|c|c|c|c|c|c|c|c|c|}
\hline
 & $1$ & $t^2$ & $t^4$ & $t^6$ & $t^8$ & $t^{10}$ & $t^{12}$ & $\dots$ \\
\hline
$HC_{0}$ & 1 & 3 & 3 & 4 & 3 & 3 & 4 & $\dots$ \\
$HC_{1}$ & 0 & 0 & 0 & 2 & 0 & 0 & 2 & $\dots$ \\
$HC_{2}$ & 0 & 0 & 0 & 1 & 0 & 0& 1 & $\dots$ \\
\hline
$\mathcal{I}_{s.t}(t)$ & 1 & 3 & 3 & 3 & 3 & 3 & 3 & $\dots$ \\
\hline
\end{tabular}
\caption{Cyclic homology group dimensions for the $\beta$-deformation}
\label{tab:beta}
\end{center}
\label{default}
\end{table}%
The operators corresponding to $\overline{HC}_{0}(\mathcal{A}) \cong \mathcal{A}/ [ \mathcal{A}, \mathcal{A} ] \cong \mathcal{A}_{cyc}$ were first determined in \cite{Berenstein:2000ux} using the techniques reviewed in section \ref{sec:cyc}.  The derivation is instructive, so we briefly recall it.
Consider an operator $\mathcal{O} = \Tr l_{1} l_{2} \dots l_{n}$, where $l_{i}$ is one of the letters $x,y$, or $z.$ Suppose that $l_1$ is an $x.$  The F-term conditions imply that
$$\mathcal{O} = \Tr l_{1} l_{2} \dots l_{n-1} l_{n} = q^{2(|z|-|y|)}  \Tr l_{n} l_{1} l_{2} \dots l_{n-1},$$
where $|x|, |y|$, and $|z|$ are the total number of $x$'s, $y$'s, and $z$'s in the operator $\mathcal{O}.$  Thus for $q$ not a primitive root of unity, $|z|-|y| =0$.
Repeating the argument with the letter $y$ or $z$, we find that the single-trace chiral primaries must have have charges $(k,0,0), (0,k,0),(0,k,0),$ or $(k,k,k)$ \cite{Berenstein:2000hy,Berenstein:2000ux,Lunin:2005jy}.  These operators exactly correspond to the graded components of $\overline{HC}_{0}.$  

For $G = SU(N)$ there are additional chiral primaries $\Tr xy, \Tr xz$ and $\Tr yz$.  This agrees with the perturbative one-loop spectrum of chiral operators found in \cite{Freedman:2005cg, Madhu:2007ew}.  For $q$ a root of unity, the cyclic homology groups jump \cite{Berenstein:2000hy}.  It would be very interesting to reproduce the spectrum of protected operators from a Kaluza-Klein analysis of the dual supergravity solution \cite{Lunin:2005jy}.
\section{Conclusion and Future Directions}
\label{sec:conclusion}
We have seen how the spectrum of protected operators in a quiver gauge theory is determined by the cyclic homology groups $\overline{HC}_{\bullet}(\mathcal{A})$ of its superpotential algebra $\mathcal{A}$.  This is a powerful new technique for analyzing this class of theories.  Using off-the-shelf mathematical results, we have easily derived new predictions for the spectrum of intensely studied theories, such as the $\beta$-deformation and more general quiver gauge theories.

For theories dual to Freund-Rubin compactifications of the form $AdS_5$ times a Sasaki-Einstein manifold, the spectrum of protected operators has already been matched by independent calculations in both the gauge and gravity theories.  However for more general compactifications it would be interesting to develop techniques to determine the spectrum of protected operators directly in supergravity.  For example, it would be a strong check of our proposal to derive the spectrum of protected operators directly from the gravity dual of the $\beta$-deformation \cite{Lunin:2005jy} and match it to the cyclic homology groups obtained in section \ref{sec:beta}.  For a massive deformation of $\mathcal{N} = 4$ supersymmetric Yang-Mills theory, the cyclic homology groups were determined in \cite{MR1121630}.  The matching of the spectrum of protected spin-2 excitations to the short Kaluza-Klein modes in the gravity solution \cite{Pilch:2000ej} will appear in a future publication.  However, it is desirable to determine the full spectrum in more general compactifications.

Passing from cyclic homology to deRham or Poisson homology is a mathematical analog of the passage from open to closed strings in AdS/CFT.  We have shown the utility of this approach by applying it to match the spectrum of protected operators in a gauge theory and its gravity dual.  This new interpretation of AdS/CFT should hopefully allow for further insight into the mathematical underpinnings of the duality.
\section*{Acknowledgements}
The author would like to thank Y. Tachikawa and J. Schmude for a previous collaboration whose results were revisited in this paper.   The author also thanks M. van den Bergh for helpful correspondence.  Finally, the author would also like to thank D. Berenstein for inspiration and lending him a copy of Loday's book on cyclic homology in his formative years.  The  work of R.~E.~is  supported in part by World Premier International Research Center Initiative
(WPI Initiative),  MEXT, Japan through the Institute for the Physics and Mathematics
of the Universe, the University of Tokyo. 
\bibliographystyle{amsplain}
\bibliography{cyclicref}
\end{document}